\documentclass[twoside,twocolumn,9pt]{article}
\usepackage{extsizes}
\usepackage[super,sort&compress,comma]{natbib}
\usepackage[left=1.5cm, right=1.5cm, top=1.785cm, bottom=2.0cm]{geometry}
\usepackage{balance}
\usepackage{mathptmx}
\usepackage{sectsty}
\usepackage{graphicx}
\usepackage{lastpage}
\usepackage[format=plain,justification=justified,singlelinecheck=false,font={stretch=1.125,small,sf},labelfont=bf,labelsep=space]{caption}
\usepackage{float}
\usepackage{fancyhdr}
\usepackage{fnpos}
\usepackage[english]{babel}
\addto{\captionsenglish}{%
  
}
\usepackage{array}
\usepackage{droidsans}
\usepackage{booktabs}
\usepackage{multirow}
\usepackage{charter}
\usepackage[T1]{fontenc}
\usepackage[usenames,dvipsnames]{xcolor}
\usepackage{setspace}
\usepackage[compact]{titlesec}
\usepackage{hyperref}

\usepackage{xfrac}
\usepackage{chemfig}
\usepackage{graphicx}
\usepackage{dcolumn}
\usepackage{bm}
\usepackage{siunitx}
\usepackage{xcolor}

\DeclareSIUnit{\calorie}{cal}

\newcommand{\mrm}{\mathrm}
\newcommand{\ket}[1]{\vert #1 \rangle}

\newcommand{\bra}[1]{\langle #1 \vert}

\newcommand{\Tbraket}[3]{\langle #1 \hspace{.10em} \vert \hspace{.10em} #2 \hspace{.10em} \vert \hspace{.10em} #3 \rangle}

\usepackage[version=4]{mhchem}

\newcommand{\simH}{\bar{H}}

\usepackage{epstopdf}

\definecolor{cream}{RGB}{222,217,201}
\begin{document}

\pagestyle{fancy}
\thispagestyle{plain}
\fancypagestyle{plain}{
\renewcommand{\headrulewidth}{0pt}
}

\makeFNbottom
\makeatletter
\renewcommand\LARGE{\@setfontsize\LARGE{15pt}{17}}
\renewcommand\Large{\@setfontsize\Large{12pt}{14}}
\renewcommand\large{\@setfontsize\large{10pt}{12}}
\renewcommand\footnotesize{\@setfontsize\footnotesize{7pt}{10}}
\renewcommand\scriptsize{\@setfontsize\scriptsize{7pt}{7}}
\makeatother

\renewcommand{\thefootnote}{\fnsymbol{footnote}}
\renewcommand\footnoterule{\vspace*{1pt}%
\color{cream}\hrule width 3.5in height 0.4pt \color{black} \vspace*{5pt}}
\setcounter{secnumdepth}{5}

\makeatletter
\renewcommand\@biblabel[1]{#1}
\renewcommand\@makefntext[1]%
{\noindent\makebox[0pt][r]{\@thefnmark\,}#1}
\makeatother
\renewcommand{\figurename}{\small{Fig.}~}
\sectionfont{\sffamily\Large}
\subsectionfont{\normalsize}
\subsubsectionfont{\bf}
\setstretch{1.125} 
\setlength{\skip\footins}{0.8cm}
\setlength{\footnotesep}{0.25cm}
\setlength{\jot}{10pt}
\titlespacing*{\section}{0pt}{4pt}{4pt}
\titlespacing*{\subsection}{0pt}{15pt}{1pt}

\fancyfoot{}
\fancyfoot[RO]{\footnotesize{\sffamily{1--\pageref{LastPage} ~\textbar  \hspace{2pt}\thepage}}}
\fancyfoot[LE]{\footnotesize{\sffamily{\thepage~\textbar\hspace{3.45cm} 1--\pageref{LastPage}}}}
\fancyhead{}
\renewcommand{\headrulewidth}{0pt}
\renewcommand{\footrulewidth}{0pt}
\setlength{\arrayrulewidth}{1pt}
\setlength{\columnsep}{6.5mm}
\setlength\bibsep{1pt}

\makeatletter
\newlength{\figrulesep}
\setlength{\figrulesep}{0.5\textfloatsep}

\newcommand{\topfigrule}{\vspace*{-1pt}%
\noindent{\color{cream}\rule[-\figrulesep]{\columnwidth}{1.5pt}} }

\newcommand{\botfigrule}{\vspace*{-2pt}%
\noindent{\color{cream}\rule[\figrulesep]{\columnwidth}{1.5pt}} }

\newcommand{\dblfigrule}{\vspace*{-1pt}%
\noindent{\color{cream}\rule[-\figrulesep]{\textwidth}{1.5pt}} }

\newcommand{\note}[1]{\textcolor{red}{[#1]}}

\makeatother

\twocolumn[
  \begin{@twocolumnfalse}
\vspace{1em}
\sffamily
\noindent\LARGE{The RAS-CCSD model, with applications for open-shell molecules.}\par
\vspace{0.3cm}

\noindent\large{Sarai Dery Folkestad$^{\ast}$}\par
\vspace{0.3cm}

\noindent\normalsize{The restricted active space coupled cluster singles and doubles (RAS-CCSD) model, and its pilot implementation is detailed. The approach offers a simple, pragmatic way to formulate a coupled cluster theory for open-shell and multireference systems. For ground state energies, errors (compared to FCI) are shown to be similar to those of unrestricted (spin-orbital) CCSD. For excited states, the errors of RAS-CCSD are generally larger, and adaptations to the methodology are needed to achieve an approach which is balanced for both ground and excited states. However, unlike unrestricted CCSD, the approach is spin-adapted and can be used to add dynamical correlation to a multideterminantal reference state.}\par

 \end{@twocolumnfalse} \vspace{0.6cm}

  ]

\renewcommand*\rmdefault{bch}\normalfont\upshape
\rmfamily
\section*{}
\vspace{-1cm}


\footnotetext{\textit{Department of Chemistry and Biomedical Science, Norwegian University of Science and Technology, 7491 Trondheim, Norway. E-mail: sarai.d.folkestad@ntnu.no}}




\sffamily{\textbf{ }}\\


\rmfamily 

\section*{Introduction}

Spin adaptation of coupled cluster theory is trivial for closed-shell species, where the cluster operator can be expressed in terms of singlet excitation operators, promoting electrons from the doubly occupied orbitals to the virtual orbitals. In the closed-shell case, all terms of the cluster operator commute. However, for high-spin open-shell systems, including dynamical correlation with coupled cluster theory in a spin-adapted manner has turned out to be a significant challenge. The cluster operator must include singlet excitation operators both creating or annihilating electrons in the singly occupied (active) orbitals. These excitation operators do not commute and the Baker--Campbell--Hausdorff (BCH) expansion of the similarity-transformed Hamiltonian no longer truncates after four nested commutators.\cite{Janssen_1991,Li_1994,Neogrady_1994,Herrmann_2020,Herrmann_2022}
The result is a severe increase in the complexity of the theory, compared to the closed-shell case.
Additionally, the cluster operator becomes more complicated in the open-shell case, and is not uniquely defined.\cite{Janssen_1991,Li_1994,Herrmann_2020}
To bypass the problems arising from non-commuting terms, several authors have advocated normal ordering of the exponential operator.\cite{Nooijen_1996,Nooijen_2001,Datta_2008,Datta_2013,Gunasekera_2024} However, the spin-adapted high-spin open-shell coupled cluster theory remains highly complicated.

Previously, we have explored entanglement coupled cluster theory,\cite{folkestad2023,folkestad2025} where the molecular system is coupled to an overall singlet with a non-interacting bath. The cluster operator is defined as in standard spin-adapted closed-shell theory, and the spin properties are enforced by a projection operator. The cluster operator is simple, but the projection operator introduces complexity in the working equations, especially if one wishes to go beyond single and double excitations in the parametrization, or want to consider total spin $S > \sfrac{1}{2}$.

Most complications of high-spin open-shell coupled cluster theory are avoided if the requirement that the wave function is an eigenfunction of $S^2$ is relaxed. Unrestricted (spin-orbital) coupled cluster theory is simple in both formulation and implementation, but the wave function is spin-contaminated, i.e., $\langle S^2\rangle \neq S(S+1)$.
While spin contamination of the unrestricted coupled cluster ground state is generally small,\cite{Stanton_1994} the equation-of-motion framework for excited states\cite{Sekino_1984,Geertsen_1989,Bartlett_2012} is highly affected by spin contamination, unless a restricted open-shell  Hartree--Fock (ROHF) reference is used. \cite{Krylov_2008}
The spin-restricted (or partially spin-adapted) approaches offer a compromise.\cite{Rittby_1988,Knowles_1993,Neogrady_1994,Szalay_1997}


For low-spin states, where the spin-adapted reference state is a configuration state function (CSF) consisting of more than a single determinant,\cite{helgaker2014} a multireference coupled cluster theory must be used. The challenges are similar to the high-spin open-shell case, and multireference coupled cluster theory is highly complicated and exists in a variety of flavours. The electronic structure community has not yet agreed on how such a theory should be formulated.\cite{Lyakh_2011,Mukherjee_1975,Lindgren_1978,Jeziorski_1981, Olipant_1991, kaldor1991,li1995, nooijen2001,Li_2003,Hanrath_2005,Kinoshita2005,Evangelista_2010,hanauer2011,datta2012multireference,gunasekera2024}

In the restricted active space coupled cluster (RAS-CC) framework, the spin-adapted cluster operator only includes excitations from doubly occupied to virtual orbitals, ensuring that the individual operators commute. The partially filled (active) orbitals are correlated in a subsequent diagonalization of the similarity-transformed Hamiltonian in a restricted active space. Hence, both ground and excited states are obtained in a spin-adapted (and spin-complete) manner, and the prohibitive complexity of spin-adapted open-shell coupled cluster theory is avoided.

The RAS-CC singles and doubles model (RAS-CCSD) has been shown to be useful for both open-shell and multireference systems. In fact, it can add dynamical correlation on top of either a restricted Hartree--Fock (closed- or open-shell) or a complete active space self consistent field (CASSCF) reference state. Furthermore, the approach is size-extensive with respect to a fixed active space.
Unlike standard coupled cluster theory, RAS-CC is convex close to ground state degeneracies, avoiding problems with multiple solutions or difficulties with convergence.\cite{Folkestad_2026_rascc}

In this paper, the RAS-CCSD equations are detailed, and its performance for the ground and excited states of open-shell molecules in different spin states is evaluated. The RAS-CCSD model is compared to full configuration interaction (FCI), spin-unrestricted coupled cluster singles and doubles (UCCSD), and to the RAS configuration interaction singles and doubles (RAS-CISD) model, showing that the ground state has the accuracy expected of a coupled cluster theory including singles and doubles in the parametrization. Excited state accuracy of RAS-CCSD is more variable, due to the linear parametrization of active-electron correlation, though it significantly improves upon RAS-CISD.

\section*{Theory}
The coupled cluster wave function\cite{bartlett2007coupled} is given by
\begin{align}
 \ket{\Psi_0} =  e^T\ket{0},
\end{align}
where the cluster operator is a sum of excitations operators $\tau_\mu$ with an associated amplitude $t_\mu$, such that
\begin{align}
 T = \sum_\mu t_\mu \tau_\mu,
\end{align}
where index $\mu$ labels the excitations.
The cluster amplitudes are determined by solving the projected coupled cluster equations:
\begin{align}
    \Omega_\mu = \Tbraket{\mu}{e^{-T}He^T}{0} = 0.\label{eq:cc}
\end{align}
In equation-of-motion coupled cluster theory,\cite{Sekino_1984,Geertsen_1989,stanton1993equation,Bartlett_2012} the similarity transformed Hamiltonian, $\simH = e^{-T}He^T$, is diagonalized to give both ground and excited states of the molecular system.
When the coupled cluster equations, eq.~\eqref{eq:cc}, are solved, the equation-of-motion Hamiltonian matrix takes the form
\begin{align}
  \boldsymbol{\simH}^{\text{CC}} =  \begin{pmatrix}
    E_{\text{CC}} & \boldsymbol{\eta}^T \\
    \boldsymbol{0} & \boldsymbol{A} + E_{\text{CC}}\boldsymbol{I}
  \end{pmatrix},
\end{align}
where the coupled cluster energy is $E_{\text{CC}} =  \Tbraket{0}{\simH}{0}$, $\eta_\mu = \Tbraket{\mu}{\simH}{0}$, and the coupled cluster Jacobian
is given by
\begin{align}
    A_{\mu\nu} =  \Tbraket{\mu}{[\simH, \tau_\nu]}{0}.
\end{align}
\begin{figure}
  \centering
  \includegraphics[width=0.35\linewidth]{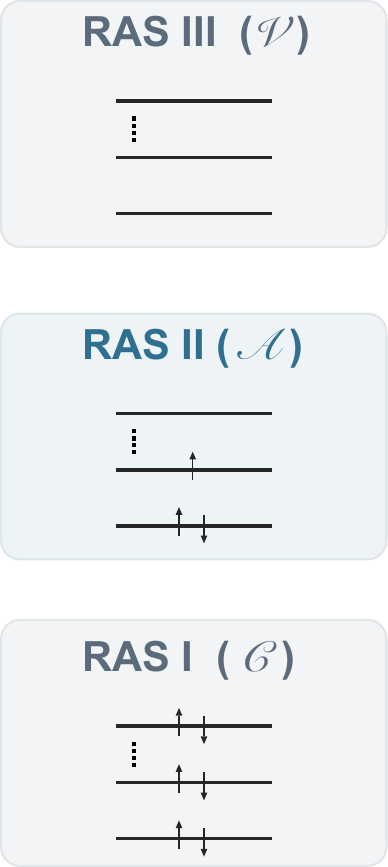}
  \caption{The restricted active space (RAS) partitioning. In the RAS-CCSD model, only two holes and particles are allowed in RAS I and III, respectively. The RAS II space can have any number of electrons consistent with the restrictions on  RAS I and III.}
  \label{fig:ras}
\end{figure}

In RAS-CC, the cluster operator is restricted to include excitations from doubly occupied orbitals to virtual orbitals, and introduce correlation for partially filled (active) orbitals through restricted active space (RAS) diagonalization of $\simH$. The RAS orbital ordering is shown in Figure \ref{fig:ras}, where the orbitals are partitioned into three sets (RAS I--III),
\begin{align}
    \mathcal{O} = \mathcal{C} \cup  \mathcal{A} \cup  \mathcal{V},
\end{align}
consisting of the doubly occupied core orbitals (RAS I, $\mathcal{C}$), the active orbitals (RAS II, $\mathcal{A}$), and the virtual orbitals (RAS III, $\mathcal{V}$).
In the following, indices $i, j, k$ denote orbitals in $\mathcal{C}$, $v, w$ denote orbitals in $\mathcal{A}$, and $a, b, c$ denote orbitals in $\mathcal{V}$.
The active orbital space (RAS II, $\mathcal{A}$) is kept minimal in RAS-CC, letting the $T$ operator include as much dynamical correlation as possible.

The first step of the RAS-CC scheme is a closed-shell spin-adapted coupled cluster calculation for the inactive orbitals, i.e., eq. \eqref{eq:cc} is solved
with $T$ restricted to excite from orbitals in $\mathcal{C}$ to orbitals in $\mathcal{V}$. For the RAS-CCSD model, the cluster operator is given by
\begin{align}
    T = \sum_{\mu\in\{\mu_1,\mu_2\}}t_{\mu}\tau_{\mu} = \sum_{ai} t^{a}_i E_{ai} + \frac{1}{2}\sum_{aibj} t^{ab}_{ij} E_{ai}E_{bj}.
\end{align}
The projection space in eq. \eqref{eq:cc}, $\{\bra{\mu}\}$, is defined by $\{\bra{\mrm{R}}\tau_\mu^\dagger\}$ where $\tau_\mu$ are the excitation operators included in $T$. Hence, the space is restricted to excited determinants where core electrons are promoted to the virtual orbitals (external excitations).
Using the Baker-Campbell-Hausdorff formula for eq. \eqref{eq:cc}, we get
\begin{align}
\begin{split}
   \Omega_\mu &= \Tbraket{\mu}{e^{-T}He^{T}}{0}  \\
   &=  \Tbraket{\mu}{\tilde{H}}{0} + \Tbraket{\mu}{[\tilde{H},T_2]}{0} +  \tfrac{1}{2}\Tbraket{\mu}{[[\tilde{H},T_2],T_2]}{0}\label{eq:ras_omega_1}
\end{split}
\end{align}
which terminates after two nested commutators since all operators in $T$ commute and the determinants in $\{\bra{\mu}\}$ are restricted to have maximally two holes and two particles in $\mathcal{C}$ and $\mathcal{V}$, respectively. In eq. \eqref{eq:ras_omega_1}, the $T_1$-transformed Hamiltonian $\tilde{H} = e^{-T_1}He^{T_1}$ has been introduced.
The $T_1$-transformed Hamiltonian operator is partitioned into an inactive, an active, and a mixed contribution,
\begin{align}
    \tilde{H} = \tilde{H}^{I} + \tilde{H}^{A} + \tilde{H}^{IA},
\end{align}
where, $\tilde{H}^{I}$ contains the operators acting only on the orbitals of $\mathcal{C}$ and $\mathcal{V}$, $\tilde{H}^{A}$ contains the operators acting only on the orbitals of $\mathcal{A}$, and $\tilde{H}^{IA}$ contain operator strings with mixed active and inactive indices.

Using this partitioning of $\tilde{H}$, and that $\Tbraket{\mu}{H^{A}}{0} = 0$ and $[H^{A},T]=0$, eq. \eqref{eq:ras_omega_1} can be written as
\begin{align}
\begin{split}
   \Omega_\mu =&  \Tbraket{\mu}{ \tilde{H}^{I}}{0} + \Tbraket{\mu}{ \tilde{H}^{IA}} {0} \\
   +& \Tbraket{\mu}{[ \tilde{H}^{I},T_2]}{0} + \Tbraket{\mu}{[ \tilde{H}^{IA},T_2]}{0}\\
   +& \Tbraket{\mu}{[[\tilde{H}^{I},T_2],T_2]}{0} + \Tbraket{\mu}{[[\tilde{H}^{IA},T_2],T_2]}{0}\\
   =& \Omega^{I}_\mu + \Tbraket{\mu}{[ \tilde{H}^{IA},T_2]}{0} + \Tbraket{\mu}{[[\tilde{H}^{IA},T_2],T_2]}{0},
\end{split}
\end{align}
where $\boldsymbol\Omega^{I}$ equals the standard closed-shell coupled cluster $\boldsymbol\Omega$-vector for the inactive orbitals. The additional terms to be evaluated are those containing $H^{IA}$. Since both $\bra{\mu}$ and $\ket{0}$ have a fixed number of electrons in the inactive (and active) orbitals, only the terms of $H^{IA}$ that conserve the number of electrons in each space can contribute. In the biorthogonal basis, the additional contributions at the CCSD level of theory are:
\begin{align}
    \Tbraket{\mu}{H^{IA}} {0} &=  \sum_{pqvw}\Tbraket{\mu}{\tilde g_{pqvw}e_{pqvw}} {0}\nonumber\\
     &= \sum_{vw}D_{vw}(\tilde g_{aivw} - \tfrac{1}{2} \tilde g_{awvi})\delta_{\mu,ai}
\\
     \Tbraket{\mu}{[H^{IA},T_2]} {0} &=  \sum_{pqvw}\Tbraket{\mu}{[(g_{pqvw}e_{pqvw}),T_2]} {0} \nonumber\\
     \begin{split}
     =\sum_{ck} \sum_{vw}&D_{vw}(\tilde g_{kcvw} - \tfrac{1}{2} \tilde g_{kwvc}) u^{ac}_{ik}\delta_{\mu,ai} \\
     + P^{ab}_{ij}&\Bigg(\sum_{c} \sum_{vw}D_{vw}(\tilde g_{bcvw} - \tfrac{1}{2}\tilde g_{bwvc}) t^{ac}_{ij}  \\
     &- \sum_{k} \sum_{vw}D_{vw}(\tilde g_{kjvw}- \tfrac{1}{2}\tilde g_{kwvj}) t^{ab}_{ik} \Bigg)\delta_{\mu, aibj}.
     \end{split}
\end{align}
Here $\tilde g_{pqrs}$ is a $T_1$-transformed electron repulsion integral, and $u^{ab}_{ij} = 2t^{ab}_{ij}  - t^{ab}_{ji}$.
Identifying $\tilde F^{A}_{pq} =  \sum_{vw}D_{vw}(\tilde g_{pqvw}- \tfrac{1}{2}\tilde g_{pwvq})$ as the active part of the inactive-inactive generalized Fock matrix,\cite{helgaker2014} these contributions can be rewritten as:
\begin{align}
    \Tbraket{\mu}{H^{IA}} {0} &=\tilde F^{A}_{ai}\delta_{\mu,ai}\\
    \begin{split}
     \Tbraket{\mu}{[H^{IA},T_2]} {0} &=
     \sum_{ck}\tilde F^{A}_{kc} u^{ac}_{ik}\delta_{\mu,ai} \\
     &+ \delta_{\mu, aibj} \cdot P^{ab}_{ij}\Bigg(\sum_{c}\tilde F^{A}_{bc}t^{ac}_{ij} - \tilde F^{A}_{kj} t^{ab}_{ik} \Bigg).
     \end{split}
\end{align}
These extra terms are most easily incorporated into the CCSD equations by using the generalized Fock matrix,\cite{helgaker2014} rather than the inactive Fock matrix, in the standard CCSD equations. Note that this is exactly the same procedure as any other frozen-orbital coupled cluster calculation, such as frozen-core or reduced-space coupled cluster calculations.
The molecular orbital density matrix for the active orbitals ($D_{vw},\;vw\in\mathcal{A}$) can be obtained from a restricted Hartree--Fock calculation (open- or closed-shell), a CASSCF calculation, or a state-averaged CASSCF calculation.

After solving the closed-shell spin-adapted coupled cluster equations, $\simH$ is diagonalized in a RAS determinant basis ($\boldsymbol{\simH}^{\text{RAS-CCSD}}$).
The RAS-CCSD determinant basis is defined by having at most two holes and two particles in $\mathcal{C}$ and $\mathcal{V}$, respectively. The electron number of $\mathcal{A}$ must therefore be restricted to the interval $[N_e^{\mathcal{A}} - 2, N_e^{\mathcal{A}} + 2]$, where $N_e^{\mathcal{A}}$ is the number of electrons in $\mathcal{A}$ in the reference state. All determinants with the correct $M_S$ and the correct total electron number $N_e$, that satisfy the given occupation restrictions, are included in the basis.
Hence, the RAS determinant basis is larger than the basis for which the coupled cluster equations are solved. Furthermore, since the RAS diagonalization is performed in a string based configuration interaction code,\cite{helgaker2014} the approach is automatically spin-complete, i.e., the relevant spin space is fully spanned for each spatial configuration.

\begin{table*}[tbh!]
\centering
\caption{Ground-state energies relative to FCI (cc-pVDZ). The FCI total energy is in \si{\hartree} and the error $\Delta = E_0-E_\mathrm{FCI}$ is given in \si{\milli\hartree}. The RAS-CISD and RAS-CCSD models are spin pure ($\langle S^2\rangle = S(S{+}1)$), and $\langle S^2\rangle$ is given for the (EOM-)UCCSD ground state.}
\label{tab:gs-fci}
\begin{tabular}{llcccccc}
\toprule
 & & & & \multicolumn{3}{c}{$\Delta\;\; [\si{\milli\hartree}]$} & \\
\cmidrule(lr){5-7}
System & $S{=}M_S$ & Ref. & $E_\mathrm{FCI}\;\; [\si{\hartree}]$ & RAS-CISD & RAS-CCSD & UCCSD & $\langle S^2\rangle$\\
\midrule
\multirow{2}{*}{BeH} & $1/2$ & ROHF & $-15.1890$ & $0.65$ & $0.62$ & $0.63$ & $0.750$\\
 & $3/2$ & SA-CASSCF & $-14.9741$ & $0.05$ & $0.05$ & $0.03$ & $3.750$\\
\addlinespace
\multirow{3}{*}{BH$_2$} & $1/2$ & ROHF & $-25.8406$ & $3.23$ & $1.52$ & $1.57$ & $0.750$\\
 & $3/2$ & ROHF & $-25.6557$ & $2.46$ & $2.27$ & $2.45$ & $3.750$\\
 & $5/2$ & ROHF & $-25.1982$ & $0.47$ & $0.47$ & $0.31$ & $8.750$\\
\addlinespace
\multirow{3}{*}{CH} & $1/2$ & SA-CASSCF & $-38.3821$ & $4.73$ & $2.49$ & $2.53$ & $0.750$\\
 & $3/2$ & ROHF & $-38.3603$ & $1.80$ & $1.54$ & $1.51$ & $3.750$\\
 & $5/2$ & ROHF & $-37.8657$ & $0.29$ & $0.29$ & $0.19$ & $8.750$\\
\addlinespace
\multirow{3}{*}{CH$_2$} & $0$ & RHF & $-39.0251$ & $9.33$ & $3.89$ & $3.89$ & $0.000$\\
 & $1$ & ROHF & $-39.0285$ & $5.06$ & $2.76$ & $2.44$ & $2.000$\\
 & $2$ & ROHF & $-38.6477$ & $4.01$ & $3.84$ & $4.53$ & $6.000$\\
\addlinespace
\multirow{3}{*}{NH} & $0$ & SA-CASSCF & $-55.0239$ & $5.14$ & $2.73$ & $3.11^{\ast}$ & $0.000$\\
 & $1$ & ROHF & $-55.0916$ & $5.10$ & $2.85$ & $2.27$ & $2.000$\\
 & $2$ & ROHF & $-54.7957$ & $1.89$ & $1.80$ & $1.45$ & $6.000$\\
\addlinespace
\multirow{3}{*}{OH} & $1/2$ & SA-CASSCF & $-75.5616$ & $6.80$ & $3.96$ & $2.24$ & $0.750$\\
 & $3/2$ & ROHF & $-75.2771$ & $3.70$ & $2.56$ & $1.63$ & $3.750$\\
 & $5/2$ & ROHF & $-74.3497$ & $1.59$ & $1.52$ & $1.09$ & $8.750$\\
\addlinespace
\multirow{3}{*}{HF} & $0$ & RHF & $-100.2306$ & $8.91$ & $2.43$ & $2.43$ & $0.000$\\
 & $1$ & SA-CASSCF & $-99.8551$ & $5.23$ & $3.74$ & $2.22^{\ast}$ & $2.000$\\
 & $2$ & ROHF & $-98.8406$ & $2.74$ & $2.10$ & $1.35$ & $6.000$\\
\midrule
\multicolumn{4}{l}{MAE $[\si{\milli\hartree}]$} & $3.66$ & $2.17$ & $1.89$ & \\
\multicolumn{4}{l}{max $|\Delta|\;\; [\si{\milli\hartree}]$} & $9.33$ & $3.96$ & $4.53$ & \\
\bottomrule
\end{tabular}
\\\footnotesize$^\ast$ ground state is obtained from a negative energy EOM-UCCSD excitation energy.
\end{table*}

The similarity-transformed Hamiltonian is non-Hermitian and properties (other than energies) are generally only obtained after both left and right eigenvectors are determined,
\begin{align}
  &\boldsymbol{\simH}^{\text{RAS-CC}} \boldsymbol{R}_k = E_k \boldsymbol{R}_k,\quad
  \boldsymbol{L}_k^T\boldsymbol{\simH}^{\text{RAS-CC}} = E_k \boldsymbol{L}_k^T,\quad
  \boldsymbol{L}_k^T \boldsymbol{R}_l = \delta_{kl},
\end{align}
though in this work, only energies are considered.
The RAS-CCSD Hamiltonian matrix elements takes the form
\begin{align}
\begin{split}
    {\simH}^{\text{RAS-CCSD}}_{\mu\nu} &= \Tbraket{\mu}{e^{-T}H e^{T}}{\nu}\\
    &= \Tbraket{\mu}{\tilde{H} + [\tilde{H},T_2] + \frac{1}{2}[[\tilde{H},T_2],T_2]}{\nu} \\
    &= {\tilde{H}}_{\mu\nu} + \Tbraket{\mu}{[\tilde{H},T_2] + \frac{1}{2}[[\tilde{H},T_2],T_2]}{\nu},
\end{split}
\end{align}
where $\tilde{H}$ is the $T_1$-transformed Hamiltonian, and $\mu$ and $\nu$ now label determinants in the restricted active space. Note that for $T_1 = 0$, ${\tilde{H}}_{\mu\nu}$ is the RAS-CISD Hamiltonian matrix.
An implementation of the linear transformation
\begin{align}
  \sigma_{K^\alpha K^\beta} =& \Tbraket{K^\alpha K^{\beta}}{\simH}{I^\alpha I^{\beta}}C_{I^\alpha I^{\beta}}\\
  \begin{split}
  =& \Tbraket{K^\alpha K^{\beta}}{\tilde H}{I^\alpha I^{\beta}}C_{I^\alpha I^{\beta}} \\
  &+ \Tbraket{K^\alpha K^{\beta}}{[\tilde H,T_2]}{I^\alpha I^{\beta}}C_{I^\alpha I^{\beta}}\\
  &+ \Tbraket{K^\alpha K^{\beta}}{[[\tilde H,T_2],T_2]}{I^\alpha I^{\beta}}C_{I^\alpha I^{\beta}}
  \end{split}
\end{align}
is needed, where, $I^\sigma$ denote a string of $\sigma$ creation operators, following Ref.~\citenum{helgaker2014}.
Note that in RAS-CCSD, we get 1) a loss of symmetry in the integrals because of the $T_1$-transformation, and 2) additional terms arising from the commutators with $T_2$, compared to a standard RAS-CI implementation. The additional terms do not lead to an increase in the scaling compared to a standard (non-symmetric) RAS-CISD calculation. The additional contributions to the equations from the single commutator are given by
\begin{align}
  [H, T_2] = & t^{ab}_{ij}\Big( h_{pa} e_{pibj} - h_{ip} e_{apbj} - g_{pbiq} e_{pjaq}
      - g_{pqja} e_{pqbi} \label{eq:T2}\\
      &+ \tfrac12 g_{pbqa} e_{pjqi}
      + \tfrac12 g_{jpiq} e_{bpaq} + g_{pqra} E_{bj}e_{pqri} - g_{pqri}E_{bj}e_{pqar}\Big),\nonumber
\end{align}
and the contribution from the doubles commutator, can be written as
\begin{equation}
  \tfrac{1}{2}[[H,T_2], T_2]\rightarrow X_{aibj}\, e_{aibj},
  \quad
  \boldsymbol{X} = \boldsymbol{X}^{(2)} + \boldsymbol{X}^{(3)} + \boldsymbol{X}^{(4)},\label{eq:T2T2_0}
\end{equation}
with
\begin{align}
  X^{(2)}_{aibj} =& \tfrac12g_{kcld}\Big(
      t^{ab}_{kl}t^{cd}_{ij}
    + t^{ac}_{ki}t^{bd}_{lj}
    + t^{ad}_{kj}t^{bc}_{li}
  \Big),\label{eq:T2T2_1}\\
    X^{(3)}_{aibj} =& - P^{ab}_{ij}\Big(t^{ab}_{ik}t^{cd}_{lj}L_{lckd}
    + t^{ac}_{ik}t^{bd}_{lj}L_{kcld}
    + t^{ac}_{ij}t^{bd}_{kl}L_{kcld}\Big),\label{eq:T2T2_2}\\
  X^{(4)}_{aibj} =& t^{ac}_{ik}\, L_{ckdl}\, t^{db}_{lj}\label{eq:T2T2_3}.
\end{align}
To arrive at eqs. \eqref{eq:T2T2_0}--\eqref{eq:T2T2_3}, the occupation restrictions in $\mathcal{C}$ and $\mathcal{V}$ have been used.

\begin{figure}
  \centering
  \includegraphics[width=\linewidth]{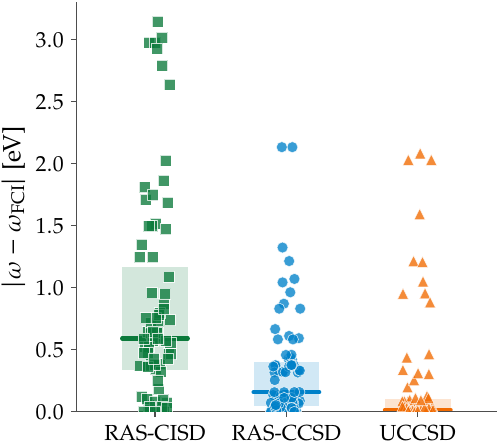}
  \caption{Errors relative to FCI of the four lowest excitation energies per $S=M_S$ of the \ce{BeH}, \ce{BH2}, \ce{CH}, \ce{CH2}, \ce{NH}, \ce{OH}, and \ce{HF} molecules.  RAS-CISD (MAE: 0.837 eV, median: 0.591 eV), RAS-CCSD (MAE: 0.327 eV, median: 0.158 eV), and UCCSD (MAE: 0.226 eV, median: 0.017 eV)\label{fig:fci-es}}
\end{figure}

\begin{figure*}[tbh!]
  \centering
  \includegraphics[width=0.9\linewidth]{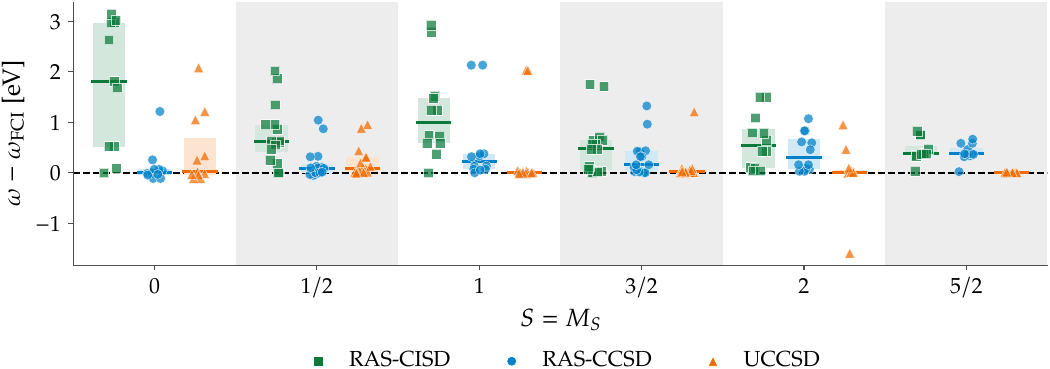}
  \caption{Excited state errors of RAS-CISD, RAS-CCSD, UCCSD compared to FCI for \ce{BeH}, \ce{BH2}, \ce{CH}, \ce{CH2}, \ce{NH}, \ce{OH}, and \ce{HF}, ordered by spin quantum number $S$.  \label{fig:fci-es-by-spin}}
\end{figure*}
All terms from $[H, T_2]$ (eq. \eqref{eq:T2}), that consist of two-body excitation operators are included by modifying the integrals of the standard RAS-CI code.
The three-body excitations are handled separately using
\begin{align*}
  & \Tbraket{K^\alpha K^{\beta}}{E_{bj}e_{pqrs}}{I^\alpha I^{\beta}}\\
  & = \sum_{J^\alpha J^{\beta}\leq 1}\Tbraket{K^\alpha K^{\beta}}{E_{bj}}{J^\alpha J^{\beta}}
  \times\Tbraket{J^\alpha J^{\beta}}{e_{pqrs}}{I^\alpha I^{\beta}}
\end{align*}
and
\begin{align}
  \Tbraket{K^\alpha K^{\beta}}{E_{bj}}{J^\alpha J^{\beta}}
    &= D^\alpha_{bj, K^\alpha, J^\alpha}\delta_{J^{\beta},K^{\beta}} + D^\beta_{bj, K^\beta, J^\beta}\delta_{J^{\alpha},K^{\alpha}}
\end{align}
where $D^\sigma_{bj, K^\sigma, J^\sigma} = \Tbraket{ K^{\sigma}}{a^\dagger_{b\sigma}a_{j\sigma}}{J^{\sigma}}$ and  $J^\alpha J^{\beta}\leq 1$ is used to denote that the determinant cannot have more than a single hole in $\mathcal{C}$ or particle in $\mathcal{V}$. This term is implemented exploiting the sparsity in $\boldsymbol{D}^\sigma$.
The $\tfrac{1}{2}[[H,T_2], T_2]$ contribution (eq. \eqref{eq:T2T2_0}) is handled separately, but also has a lower cost than the standard RAS-CISD equations.
The current pilot implementation is built using a single integer for the $\alpha$ and $\beta$ strings of each orbital space. Hence, with 64-bit integers, the current limitation is 64 orbitals in each of $\mathcal{C}$, $\mathcal{A}$, and $\mathcal{V}$. The number of orbitals can be increased in a straightforward manner by using several integers per space, or, alternatively with a generalized active space implementation. However, code optimizations are needed before such extensions are useful.

\section*{Results and Discussion}

The RAS-CCSD code is implemented in a development version of the e$^{T}$ program.\cite{et-v1.0-jcp-2020,et-v2.0-jcp-2026}
The RHF and ROHF reference calculations are performed in e$^{T}$. However, when a CASSCF or SA-CASSCF reference state is used, the orbitals and active orbital density ($D_{vw}$, $v,w\in\mathcal{A}$) are obtained from a separate PySCF\cite{sun2018pyscf,sun2020recent,sun2026python} calculation. The following molecules are considered: \ce{BeH}, \ce{BH2}, \ce{CH}, \ce{CH2}, \ce{NH}, \ce{OH}, \ce{HF}, \ce{CH3}, \ce{B2}, \ce{C2}, \ce{H2O}, \ce{C2H}, \ce{BeO}, \ce{CN}, \ce{BO}, \ce{HCO}, \ce{NO}, \ce{CF}, \ce{O2}, \ce{HO2}, \ce{NF}, \ce{NCO}, \ce{OF}, \ce{NO2}, and \ce{O3}. The cc-pVDZ basis is used throughout, and the same geometry is used for all spin states of a molecule (Table \ref{tab:geom}). An SA-CASSCF reference is used when RHF or ROHF yields a symmetry-broken reference, averaging over the degenerate components. A state-specific CASSCF reference is used when the largest (spin-adapted) $T_2$ amplitude of the RAS-CCSD calculation with an RHF or ROHF reference exceeds 0.2 (\ce{B2} $S=1$, \ce{C2} and \ce{O3} $S=0$). Details of the active spaces and reference state types of the calculations are given in Table \ref{tab:details} in the Appendix.
The Q-Chem program\cite{Epifanovsky2015} is used for all UCCSD calculations. An RHF or ROHF state is always used as a reference in UCCSD calculations. The FCI calculations are performed with e$^{T}$.

\begin{figure}[tbh!]
  \centering
  \includegraphics[width=\linewidth]{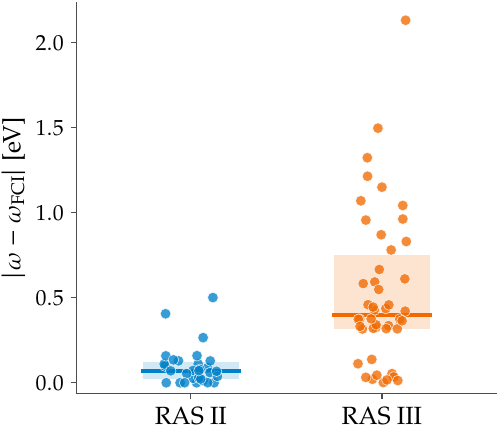}
  \caption{Excited state errors of RAS-CCSD compared to FCI for \ce{BeH}, \ce{BH2}, \ce{CH}, \ce{CH2}, \ce{NH}, \ce{OH}, and \ce{HF} categorized according to whether the dominant excitation is into, or within, RAS II ($\mathcal{A}$) or into RAS III ($\mathcal{V}$). \label{fig:ras-es-classification}}
\end{figure}

In Table \ref{tab:gs-fci}, the RAS-CCSD, RAS-CISD, and UCCSD ground state energies are compared to FCI, for a set of small molecules (\ce{BeH}, \ce{BH2}, \ce{CH}, \ce{CH2}, \ce{NH}, \ce{OH}, and \ce{HF}) and different spin quantum numbers.  The RAS-CISD and RAS-CCSD models are eigenfunctions of $S^2$, and the error in $\langle S^2\rangle$ is small for the UCCSD ground states. Note that the reported UCCSD ground state energy is the lowest energy obtained in an EOM-UCCSD calculation. For instance, for \ce{NH} ($S=0$) and \ce{HF} ($S=1$) the ground state is obtained as an equation-of-motion state with negative excitation energy. As is expected, the lack of size-extensivity in RAS-CISD results in errors that generally grow with system size. RAS-CCSD and UCCSD can be said to give energies of approximately the same quality: UCCSD having a slightly lower mean absolute error and higher maximal absolute error.

\begin{table*}[tbh!]\centering
\small
\caption{Ground-state energies, $E_0$, of UCCSD, and the deviations $\Delta = E_0-E_0^{\text{UCCSD}}$ of RAS-CISD and RAS-CCSD. $\langle S^2\rangle$ is that of the (EOM-)UCCSD ground state. Reference types and active spaces for the RAS-CISD and RAS-CCSD calculations are listed in Table \ref{tab:details}. $\Delta_\mathrm{ref}=E_\mathrm{ref}^{\text{RAS-CCSD}}-E_\mathrm{ref}^{\text{UCCSD}}$ is the difference of the reference energies.}
\label{tab:gs-secondrow}
\begin{tabular}{ll l c c c c c}
\toprule
 & & & & & \multicolumn{2}{c}{$\Delta = E_0-E_0^{\text{UCCSD}}\;\;[\si{\milli\hartree}]$} & \\
\cmidrule(lr){6-7}
System & $S{=}M_S$ & Ref. & $\Delta_\mathrm{ref}\;\;[\si{\milli\hartree}]$ & $E_0^{\text{UCCSD}}\;\;[\si{\hartree}]$ & RAS-CISD & RAS-CCSD & $\langle S^2\rangle$\\
\midrule
\multirow{3}{*}{CH$_3$} & $1/2$ & ROHF & $0.00$ & $-39.7157$ & $+5.41$ & $-0.07$ & $0.750$\\
                        & $3/2$ & SA-CASSCF & $+25.28$ & $-39.2950$ & $+6.95$ & $+6.15$ & $3.750$\\
                        & $5/2$ & SA-CASSCF & $+9.13$ & $-38.8325$ & $-2.07$ & $-2.21$ & $8.750$\\
\addlinespace
\multirow{3}{*}{B$_2$} & $0$ & SA-CASSCF & $-31.86$ & $-49.2329^\ast$ & $-1.55$ & $-5.65$ & $0.000$\\
                       & $1$ & CASSCF & $-45.35$ & $-49.2654^\ast$ & $-2.71$ & $-3.64$ & $2.469$\\
                       & $2$ & ROHF & $0.00$ & $-49.2616$ & $+0.42$ & $-0.03$ & $6.001$\\
 \addlinespace
 \multirow{3}{*}{C$_2$} & $0$ & CASSCF & $-86.76$ & $-75.7023$ & $-0.15$ & $-1.86$ & $0.000$\\
                        & $1$ & ROHF & $+9.43$ & $-75.7039$ & $+11.96$ & $+0.73$ & $2.001$\\
                        & $2$ & ROHF & $+11.75$ & $-75.5097^\ast$ & $+7.52$ & $+2.43$ & $6.012$\\
\addlinespace
 \multirow{3}{*}{H$_2$O} & $0$ & RHF & $0.00$ & $-76.2381$ & $+8.28$ & $0.00$ & $0.000$\\
                         & $1$ & ROHF & $-77.40$ & $-75.9680^\ast$ & $+5.15$ & $+0.27$ & $2.000$\\
                         & $2$ & ROHF & $0.00$ & $-75.5130$ & $+2.22$ & $+1.22$ & $6.000$\\
 \addlinespace
\multirow{2}{*}{C$_2$H} & $1/2$ & ROHF & $0.00$ & $-76.3929$ & $+18.21$ & $-0.32$ & $0.752$\\
                        & $5/2$ & ROHF & $0.00$ & $-76.0215$ & $+4.84$ & $+2.48$ & $8.750$\\
\addlinespace
\multirow{3}{*}{BeO} & $0$ & RHF & $0.00$ & $-89.6513$ & $+14.20$ & $0.00$ & $0.000$\\
                     & $1$ & SA-CASSCF & $-40.64$ & $-89.6156^\ast$ & $+4.41$ & $+1.48$ & $2.000$\\
                     & $2$ & SA-CASSCF & $+1.22$ & $-89.3361$ & $+2.43$ & $+0.93$ & $6.000$\\
\addlinespace
\multirow{3}{*}{CN} & $1/2$ & ROHF & $0.00$ & $-92.4604$ & $+26.85$ & $-0.48$ & $0.752$\\
                    & $3/2$ & SA-CASSCF & $+12.45$ & $-92.2957$ & $+27.37$ & $-8.55$ & $3.751$\\
                    & $5/2$ & ROHF & $-103.57$ & $-92.1524^\ast$ & $+2.98$ & $+1.06$ & $8.750$\\
\addlinespace
\multirow{2}{*}{BO} & $1/2$ & ROHF & $0.00$ & $-99.7749$ & $+14.10$ & $-0.54$ & $0.751$\\
                    & $5/2$ & ROHF & $-91.78$ & $-99.2245^\ast$ & $+2.74$ & $+1.17$ & $8.750$\\
\addlinespace
\multirow{2}{*}{HCO} & $1/2$ & ROHF & $0.00$ & $-113.5690$ & $+21.33$ & $+0.17$ & $0.750$\\
                     & $3/2$ & ROHF & $0.00$ & $-113.3386$ & $+12.57$ & $+2.87$ & $3.750$\\
\addlinespace
\multirow{3}{*}{NO} & $1/2$ & SA-CASSCF & $+1.75$ & $-129.5901$ & $+21.64$ & $+5.04$ & $0.750$\\
                    & $3/2$ & ROHF & $0.00$ & $-129.3611$ & $+15.21$ & $+4.22$ & $3.751$\\
                    & $5/2$ & SA-CASSCF & $-182.98$ & $-128.7461^\ast$ & $+5.25$ & $+3.30$ & $8.751$\\
 \addlinespace
 \multirow{2}{*}{CF} & $1/2$ & SA-CASSCF & $+0.69$ & $-137.4711$ & $+16.99$ & $+0.78$ & $0.750$\\
                     & $3/2$ & ROHF & $0.00$ & $-137.3450$ & $+10.28$ & $+0.30$ & $3.750$\\
\addlinespace
\multirow{3}{*}{O$_2$} & $0$ & SA-CASSCF & $-16.55$ & $-149.9115^\ast$ & $+31.04$ & $+6.30$ & $0.000$\\
                       & $1$ & ROHF & $+27.37$ & $-149.9500$ & $+30.59$ & $+8.32$ & $2.001$\\
                       & $2$ & SA-CASSCF & $-87.38$ & $-149.6591^\ast$ & $+4.62$ & $+0.63$ & $6.002$\\
 \addlinespace
\multirow{3}{*}{HO$_2$} & $1/2$ & ROHF & $-21.47$ & $-150.5548^\ast$ & $+28.84$ & $+1.89$ & $0.750$\\
                        & $3/2$ & ROHF & $0.00$ & $-150.3198$ & $+27.57$ & $+9.22$ & $3.751$\\
                        & $5/2$ & ROHF & $-69.63$ & $-149.9602^\ast$ & $+10.39$ & $+3.94$ & $8.750$\\
\addlinespace
\multirow{3}{*}{NF} & $0$ & SA-CASSCF & $-30.44$ & $-154.0392^\ast$ & $+17.14$ & $-0.05$ & $0.000$\\
                    & $1$ & ROHF & $0.00$ & $-154.1009$ & $+17.76$ & $+1.19$ & $2.000$\\
                    & $2$ & ROHF & $0.00$ & $-153.7044$ & $+12.00$ & $+2.94$ & $6.000$\\
\addlinespace\multirow{3}{*}{NCO} & $1/2$ & SA-CASSCF & $+2.63$ & $-167.5837$ & $+45.55$ & $+7.50$ & $0.752$\\
                                  & $3/2$ & ROHF & $0.00$ & $-167.3269^\ast$ & $+39.47$ & $+8.17$ & $3.751$\\
                                  & $5/2$ & ROHF & $+54.85$ & $-166.9803$ & $+40.04$ & $+20.90$ & $8.751$\\
\addlinespace
\multirow{2}{*}{OF} & $1/2$ & SA-CASSCF & $-224.63$ & $-174.4936^\ast$ & $+22.99$ & $+2.75$ & $0.755$\\
                    & $3/2$ & ROHF & $-149.85$ & $-174.3500^\ast$ & $+20.36$ & $+4.95$ & $3.751$\\
\addlinespace
\multirow{3}{*}{NO$_2$} & $1/2$ & ROHF & $0.00$ & $-204.5855$ & $+56.23$ & $+0.33$ & $0.751$\\
                        & $3/2$ & ROHF & $0.00$ & $-204.4292$ & $+44.54$ & $+10.74$ & $3.751$\\
                        & $5/2$ & ROHF & $0.00$ & $-203.9646$ & $+38.10$ & $+17.33$ & $8.751$\\
\addlinespace
\multirow{3}{*}{O$_3$} & $0$ & CASSCF & $-95.01$ & $-224.8817$ & $+34.97$ & $-16.82$ & $0.000$\\
                       & $1$ & ROHF & $0.00$ & $-224.8327^\ast$ & $+73.17$ & $+18.92$ & $2.003$\\
                       & $2$ & ROHF & $+0.37$ & $-224.5742$ & $+48.94$ & $+16.71$ & $6.001$\\
\bottomrule
\end{tabular}\\
\footnotesize $^\ast$ ground state is obtained from a negative energy EOM-UCCSD excitation energy.
\end{table*}
In Figure \ref{fig:fci-es}, the error compared to FCI of the four lowest excitation energies of the same small molecules and different spin quantum numbers are shown for RAS-CISD, RAS-CCSD, and UCCSD. All three implementations yield excited states of a given $M_S$ and the appropriate states with $S=M_S$ must be extracted. The UCCSD states are assigned quantum number $S$ by calculating the minimal difference to the theoretical value $|\langle S^2\rangle -S(S+1)|$. Each of the four lowest FCI states is then paired with the closest state of the same $S$. Note that computing $\langle S^2\rangle$ more than doubles the computational cost of the UCCSD calculation.
While the RAS-CCSD model improves on the RAS-CISD model, it performs worse than UCCSD.

In Figure \ref{fig:fci-es-by-spin}, the excitation energies are categorised into spin quantum number $S$. For $S=0$ and $S=\sfrac{1}{2}$, the RAS-CCSD model performs as well as or better than UCCSD (MAE of \SI{0.18}{\eV} vs \SI{0.48}{\eV} for $S=0$, and \SI{0.20}{\eV} vs \SI{0.22}{\eV} for $S=\sfrac{1}{2}$), because the approach can build dynamical correlation on top of a multideterminantal reference. For \ce{CH} and \ce{OH} ($S=\sfrac{1}{2}$) and \ce{NH} ($S=0$), the RHF or ROHF references used for UCCSD are symmetry broken, while the SA-CASSCF reference used in RAS-CCSD is not.
For closed-shell singlets, RAS-CCSD and UCCSD both become the standard closed-shell EOM-CCSD approach. Hence, the RAS models are, when the appropriate reference is used, adapted for both spatial and spin symmetry.

As $S$ increases, the quality of RAS-CCSD compared to UCCSD deteriorates. As expected, for high $S=M_S$, RAS-CCSD becomes similar to RAS-CISD for these small molecular systems. This is because there are only a few doubly occupied orbitals ($\mathcal{C}$), and the benefit of the similarity transformation by $T$ becomes small. The quality of RAS-CCSD compared to RAS-CISD depends on the number of inactive electrons, and is expected to improve as molecular system size increases. Considering the dominant excitation of each excited state, it is evident that the excitations into RAS III are of significantly poorer quality, compared to those into, or within, the active space (RAS II), see Figure \ref{fig:ras-es-classification}.

Table \ref{tab:gs-secondrow}, shows the comparison of the RAS-CCSD and UCCSD ground state energies of the molecular systems \ce{CH3}, \ce{B2}, \ce{C2}, \ce{H2O}, \ce{C2H}, \ce{BeO}, \ce{CN}, \ce{BO}, \ce{HCO}, \ce{NO}, \ce{CF}, \ce{O2}, \ce{HO2}, \ce{NF}, \ce{NCO}, \ce{OF}, \ce{NO2} and \ce{O3}. Many of the systems are linear (or planar), and will have single $\pi$-level holes or particles, making a SA-CASSCF reference necessary to avoid point-group symmetry breaking (see Tables \ref{tab:gs-secondrow} and \ref{tab:details}), and some systems are of significant multireference character, such as \ce{B2}, \ce{C2}, and \ce{O3}. The former case will be symmetry-broken in UCCSD, which always builds correlation on top of an RHF or ROHF reference. Spin-contamination of the EOM-UCCSD ground state is generally small, only \ce{B2} shows significant ground state spin contamination. The deviation between the two models is generally larger for the largest systems in the test set, which can be understood as a consequence of the linear inclusion of the dynamical correlation between the active and inactive electrons. However, this is an effect of the small size of the molecular systems under consideration, since dynamical correlation is highly localizable\cite{pulay1983,saebo1993} and the approach is size-extensive (with respect to non-interacting subsystems) for a fixed active space.\cite{Folkestad_2026_rascc} An energy lowering in UCCSD compared to RAS-CCSD can also result from an increased flexibility in the parametrization due to symmetry breaking (space and spin) or from differences in the reference state, Table \ref{tab:gs-secondrow} also shows the difference in reference energy.

\begin{figure}
  \centering
  \includegraphics[width=\linewidth]{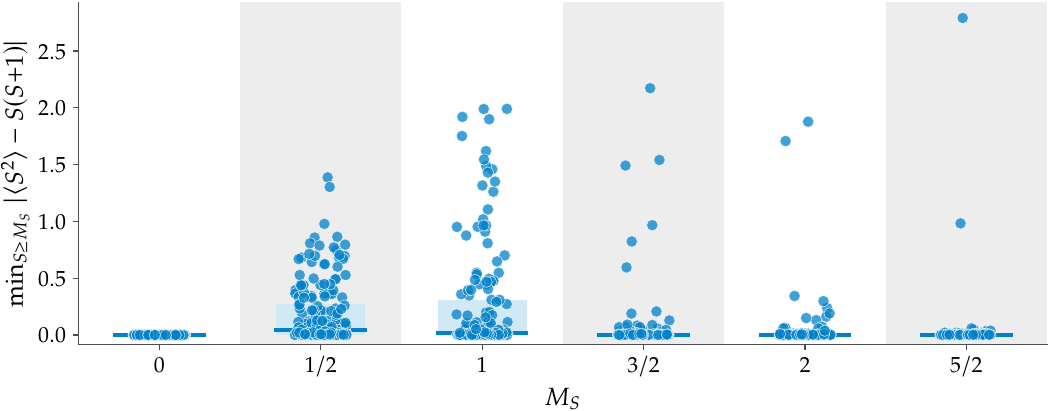}
  \caption{Spin-contamination in UCCSD ground and excited states by quantum number $M_S$. Spin-contamination is calculated as $\min_{S\geq M_S}(|\langle S^2\rangle -S(S+1)|).$ \label{fig:uccsd-es-contamination}}
\end{figure}

\begin{figure}
  \centering
  \includegraphics[width=\linewidth]{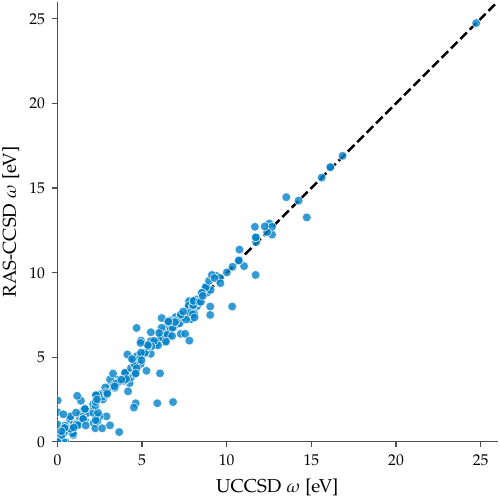}
  \caption{The four lowest RAS-CCSD and UCCSD excitation energies in $\si{\eV}$ compared for all molecules. States are filtered by spin ($S=M_S$) and paired by energy.\label{fig:ras-vs-uccsd}}
\end{figure}

Despite there being little spin contamination in the UCCSD ground state, there are many spin-contaminated excited states. In Figure \ref{fig:uccsd-es-contamination}, the spin contamination,
calculated as $\min_{S\geq M_S}(|\langle S^2\rangle -S(S+1)|)$, is shown for each of the UCCSD calculations, organized by the specified $M_S$. The approach is spin adapted for $S = 0$, displays a significant spread in $S$ when $M_S = \sfrac{1}{2}$ or $M_S = 1$. Despite there being several outliers, the spread in spin contamination errors decreases as $M_S$ increases.
In Figure \ref{fig:ras-vs-uccsd}, the lowest four $M_S = S$ RAS-CCSD and UCCSD excitation energies of the whole set of molecules are compared. Each of the four lowest RAS-CCSD states is paired with the closest UCCSD state of the same $S$. Of the 276 pairs, the mean absolute difference is $\SI{0.408}{\eV}$ and the median is $\SI{0.216}{\eV}$, 34\% of pairs are within $\SI{0.1}{\eV}$ of each other, and 59\% are within $\SI{0.3}{\eV}$.

\section*{Summary and concluding remarks}
In this paper, the RAS-CCSD implementation has been detailed and tested for a range of small molecular systems. Comparisons were made to FCI and UCCSD. Comparing to FCI, the ground state performs similarly to UCCSD. For excited states dominated by excitations into RAS III, the error is, however, significantly larger. The main benefit of the new approach is that it can add dynamical correlation on top of a multideterminantal reference, avoiding issues with spatial symmetry breaking of the reference in cases of partially filled degenerate orbitals. Furthermore, the approach is spin-adapted and spin-complete.

\section*{Conflicts of interest}
The author declares no conflict of interest.

\section*{Data availability}
The molecular geometries are given in Table \ref{tab:geom}.

\section*{Acknowledgments}
I thank Henrik Koch for discussions and computational resources.

\appendix
\section*{Appendix: Active spaces, reference states, and geometries}
Table \ref{tab:details} shows the reference state and active space used in the RAS-CCSD calculations, and Table \ref{tab:geom} the molecular geometries.
\setcounter{table}{0}
\renewcommand{\thetable}{A\arabic{table}}
\renewcommand{\theHtable}{A\arabic{table}}

\begin{table*}[htb!]\centering
\caption{Open-shell test set (cc-pVDZ). For each spin state $S{=}M_S$, the reference type and the RAS active space $\{n_\alpha,n_\beta,n_\mathrm{orb}\}$ is given.}
\label{tab:details}
\small
\setlength{\tabcolsep}{5pt}
\begin{tabular}[t]{llcc}
\toprule
System & $S{=}M_S$ & Ref. & Active $\{n_\alpha,n_\beta,n_\mathrm{orb}\}$\\
\midrule
\multirow{2}{*}{BeH} & $1/2$ & ROHF & $\{1,0,1\}$\\
 & $3/2$ & SA-CASSCF & $\{3,0,4\}$\\
\addlinespace
\multirow{3}{*}{BH$_2$} & $1/2$ & ROHF & $\{1,0,1\}$\\
 & $3/2$ & ROHF & $\{3,0,3\}$\\
 & $5/2$ & ROHF & $\{5,0,5\}$\\
\addlinespace
\multirow{3}{*}{CH} & $1/2$ & SA-CASSCF & $\{1,0,2\}$\\
 & $3/2$ & ROHF & $\{3,0,3\}$\\
 & $5/2$ & ROHF & $\{5,0,5\}$\\
\addlinespace
\multirow{3}{*}{CH$_2$} & $0$ & RHF & $\{0,0,0\}$\\
 & $1$ & ROHF & $\{2,0,2\}$\\
 & $2$ & ROHF & $\{4,0,4\}$\\
\addlinespace
\multirow{3}{*}{NH} & $0$ & SA-CASSCF & $\{1,1,2\}$\\
 & $1$ & ROHF & $\{2,0,2\}$\\
 & $2$ & ROHF & $\{4,0,4\}$\\
\addlinespace
\multirow{3}{*}{OH} & $1/2$ & SA-CASSCF & $\{2,1,2\}$\\
 & $3/2$ & ROHF & $\{3,0,3\}$\\
 & $5/2$ & ROHF & $\{5,0,5\}$\\
\addlinespace
\multirow{3}{*}{HF} & $0$ & RHF & $\{0,0,0\}$\\
 & $1$ & SA-CASSCF & $\{3,1,3\}$\\
 & $2$ & ROHF & $\{4,0,4\}$\\
\addlinespace
\multirow{3}{*}{CH$_3$} & $1/2$ & ROHF & $\{1,0,1\}$\\
 & $3/2$ & SA-CASSCF & $\{4,1,4\}$\\
 & $5/2$ & SA-CASSCF & $\{5,0,6\}$\\
\addlinespace
\multirow{3}{*}{B$_2$} & $0$ & SA-CASSCF & $\{1,1,3\}$\\
 & $1$ & CASSCF & $\{3,1,4\}$\\
 & $2$ & ROHF & $\{4,0,4\}$\\
\addlinespace
\multirow{3}{*}{C$_2$} & $0$ & CASSCF & $\{3,3,4\}$\\
 & $1$ & ROHF & $\{2,0,2\}$\\
 & $2$ & ROHF & $\{4,0,4\}$\\
\addlinespace
\multirow{3}{*}{H$_2$O} & $0$ & RHF & $\{0,0,0\}$\\
 & $1$ & ROHF & $\{2,0,2\}$\\
 & $2$ & ROHF & $\{4,0,4\}$\\
\addlinespace
\multirow{2}{*}{C$_2$H} & $1/2$ & ROHF & $\{1,0,1\}$\\
 & $5/2$ & ROHF & $\{5,0,5\}$\\
\bottomrule
\end{tabular}
\hspace{1cm}
\begin{tabular}[t]{llcc}
\toprule
System & $S{=}M_S$ & Ref. & Active $\{n_\alpha,n_\beta,n_\mathrm{orb}\}$\\
\midrule
\multirow{3}{*}{BeO} & $0$ & RHF & $\{0,0,0\}$\\
 & $1$ & SA-CASSCF & $\{3,1,3\}$\\
 & $2$ & SA-CASSCF & $\{4,0,5\}$\\
\addlinespace
\multirow{3}{*}{CN} & $1/2$ & ROHF & $\{1,0,1\}$\\
 & $3/2$ & SA-CASSCF & $\{3,0,4\}$\\
 & $5/2$ & ROHF & $\{5,0,5\}$\\
\addlinespace
\multirow{2}{*}{BO} & $1/2$ & ROHF & $\{1,0,1\}$\\
 & $5/2$ & ROHF & $\{5,0,5\}$\\
\addlinespace
\multirow{2}{*}{HCO} & $1/2$ & ROHF & $\{1,0,1\}$\\
 & $3/2$ & ROHF & $\{3,0,3\}$\\
\addlinespace
\multirow{3}{*}{NO} & $1/2$ & SA-CASSCF & $\{1,0,2\}$\\
 & $3/2$ & ROHF & $\{3,0,3\}$\\
 & $5/2$ & SA-CASSCF & $\{6,1,6\}$\\
\addlinespace
\multirow{2}{*}{CF} & $1/2$ & SA-CASSCF & $\{1,0,2\}$\\
 & $3/2$ & ROHF & $\{3,0,3\}$\\
\addlinespace
\multirow{3}{*}{O$_2$} & $0$ & SA-CASSCF & $\{1,1,2\}$\\
 & $1$ & ROHF & $\{2,0,2\}$\\
 & $2$ & SA-CASSCF & $\{5,1,5\}$\\
\addlinespace
\multirow{3}{*}{HO$_2$} & $1/2$ & ROHF & $\{1,0,1\}$\\
 & $3/2$ & ROHF & $\{3,0,3\}$\\
 & $5/2$ & ROHF & $\{5,0,5\}$\\
\addlinespace
\multirow{3}{*}{NF} & $0$ & SA-CASSCF & $\{1,1,2\}$\\
 & $1$ & ROHF & $\{2,0,2\}$\\
 & $2$ & ROHF & $\{4,0,4\}$\\
\addlinespace
\multirow{3}{*}{NCO} & $1/2$ & SA-CASSCF & $\{2,1,2\}$\\
 & $3/2$ & ROHF & $\{3,0,3\}$\\
 & $5/2$ & ROHF & $\{5,0,5\}$\\
\addlinespace
\multirow{2}{*}{OF} & $1/2$ & SA-CASSCF & $\{2,1,2\}$\\
 & $3/2$ & ROHF & $\{3,0,3\}$\\
\addlinespace
\multirow{3}{*}{NO$_2$} & $1/2$ &  ROHF  & $\{1,0,1\}$\\
 & $3/2$ & ROHF & $\{3,0,3\}$\\
 & $5/2$ & ROHF & $\{5,0,5\}$\\
\addlinespace
\multirow{3}{*}{O$_3$} & $0$ & CASSCF & $\{1,1,2\}$\\
 & $1$ & ROHF & $\{2,0,2\}$\\
 & $2$ & ROHF & $\{4,0,4\}$\\
\bottomrule
\end{tabular}
\end{table*}

\begin{table*}[htb!]\centering
\caption{Molecular geometries.}
\label{tab:geom}
\small
\begin{tabular}[t]{ll}
\toprule
System & Geometry\\
\midrule
BeH & $R_\mathrm{BeH}=\SI{1.3260}{\angstrom}$\\
BH$_2$ & $R_\mathrm{BH}=\SI{1.1811}{\angstrom}$, $\angle_\mathrm{HBH}=\ang{120.04}$\\
CH & $R_\mathrm{CH}=\SI{1.1389}{\angstrom}$\\
CH$_2$ & $R_\mathrm{CH}=\SI{1.1166}{\angstrom}$, $\angle_\mathrm{HCH}=\ang{102.40}$\\
NH & $R_\mathrm{NH}=\SI{1.1163}{\angstrom}$\\
OH & $R_\mathrm{OH}=\SI{0.9697}{\angstrom}$\\
HF & $R_\mathrm{HF}=\SI{0.9170}{\angstrom}$\\
CH$_3$ & $R_\mathrm{CH}=\SI{1.0825}{\angstrom}$, $\angle_\mathrm{HCH}=\ang{120.00}$\\
B$_2$ & $R_\mathrm{BB}=\SI{1.5900}{\angstrom}$\\
C$_2$ & $R_\mathrm{CC}=\SI{1.2425}{\angstrom}$\\
H$_2$O & $R_\mathrm{OH}=\SI{0.9755}{\angstrom}$, $\angle_\mathrm{HOH}=\ang{110.57}$\\
C$_2$H & $R_\mathrm{CH}=\SI{1.0467}{\angstrom}$, $R_\mathrm{CC}=\SI{1.2165}{\angstrom}$, $\angle_\mathrm{HCC}=\ang{180.00}$\\
BeO & $R_\mathrm{BeO}=\SI{1.3309}{\angstrom}$\\
\bottomrule
\end{tabular}
\hspace{0.5cm}
\begin{tabular}[t]{ll}
\toprule
System & Geometry\\
\midrule
CN & $R_\mathrm{CN}=\SI{1.3000}{\angstrom}$\\
BO & $R_\mathrm{BO}=\SI{1.2049}{\angstrom}$\\
HCO & $R_\mathrm{CO}=\SI{1.1753}{\angstrom}$, $R_\mathrm{CH}=\SI{1.1191}{\angstrom}$, $\angle_\mathrm{OCH}=\ang{124.95}$\\
NO & $R_\mathrm{NO}=\SI{1.1508}{\angstrom}$\\
CF & $R_\mathrm{CF}=\SI{1.2718}{\angstrom}$\\
O$_2$ & $R_\mathrm{OO}=\SI{1.3780}{\angstrom}$\\
HO$_2$ & $R_\mathrm{OO}=\SI{1.3305}{\angstrom}$, $R_\mathrm{OH}=\SI{0.9707}{\angstrom}$, $\angle_\mathrm{OOH}=\ang{104.30}$\\
NF & $R_\mathrm{NF}=\SI{1.3170}{\angstrom}$\\
NCO & $R_\mathrm{NC}=\SI{1.2049}{\angstrom}$, $R_\mathrm{CO}=\SI{1.1800}{\angstrom}$, $\angle_\mathrm{NCO}=\ang{180.00}$\\
OF & $R_\mathrm{OF}=\SI{1.3541}{\angstrom}$\\
NO$_2$ & $R_\mathrm{NO}=\SI{1.1934}{\angstrom}$, $\angle_\mathrm{ONO}=\ang{134.10}$\\
O$_3$ & $R_\mathrm{OO}=\SI{1.2717}{\angstrom}$, $\angle_\mathrm{OOO}=\ang{116.78}$\\
\bottomrule
\end{tabular}
\end{table*}

\bibliography{old/manuscript} 
\bibliographystyle{rsc} 

\end{document}